\documentclass[sigconf , nonacm]{acmart}
\usepackage{algorithm}
\usepackage{algpseudocode}
\usepackage{enumitem}
\AtBeginDocument{%
  }

\setcopyright{acmlicensed}
\copyrightyear{2018}
\acmYear{2018}
\acmDOI{XXXXXXX.XXXXXXX}
\acmConference[Conference acronym 'XX]{Make sure to enter the correct
  conference title from your rights confirmation email}{June 03--05,
  2018}{Woodstock, NY}
\acmISBN{978-1-4503-XXXX-X/2018/06}

\begin{document}

\title{PIM-FW: Hardware-Software Co-Design of All-pairs Shortest Paths in DRAM}


%

\author{Tsung-Han Lu}
\authornotemark[1]
\email{tsl012@ucsd.edu}
\affiliation{%
  \institution{University of California, San Diego}
  \city{San Diego}
  \state{California}
  \country{USA}
}

\author{Zheyu Li}
\authornotemark[2]
\email{zhl178@ucsd.edu}
\affiliation{%
  \institution{University of California, San Diego}
  \city{San Diego}
  \state{California}
  \country{USA}
}

\author{Minxuan Zhou}
\authornotemark[3]
\email{mzhou26@iit.edu}
\affiliation{%
  \institution{Illinois Institute of Technology}
  \city{Chicago}
  \state{Illinois}
  \country{USA}
}

\author{John Hsu}
\authornotemark[4]
\email{joh048@ucsd.edu}
\affiliation{%
  \institution{University of California, San Diego}
  \city{San Diego}
  \state{California}
  \country{USA}
}

\author{Tajana Rosing}
\authornotemark[5]
\email{tajana@ucsd.edu}
\affiliation{%
  \institution{University of California, San Diego}
  \city{San Diego}
  \state{California}
  \country{USA}
}









\begin{abstract}
All‐pairs shortest paths (APSP) is a fundamental algorithm used for routing, logistics, and network analysis, but the cubic time complexity and heavy data movement of the canonical Floyd–Warshall (FW) algorithm severely limits its scalability on conventional CPUs or GPUs. In this paper, we propose PIM-FW, a novel co-designed hardware architecture and dataflow that leverages processing in and near memory architecture  designed to accelerate blocked FW algorithm on an HBM3 stack.  To enable fine-grained parallelism, we propose a massively parallel array of specialized bit-serial bank PE and channel PE designed to accelerate the core min-plus operations. Our novel dataflow complements this hardware, employing an interleaved mapping policy for superior load balancing and hybrid in and near memory computing model for efficient computation and reduction. The novel in-bank computing approach allows all distance updates to be performed and stored in memory bank, a key contribution is that eliminates the data-movement bottleneck inherent in GPU-based approaches. We implement a full software and hardware co‐design using a cycle-accurate simulator to simulate an 8‐channel, 4‐Hi HBM3 PIM stack on real road‐network traces. Experimental results show that, for a \(8,192\times8,192\) graph, PIM‐FW achieves a 18.7$\times$ speedup in end‐to‐end execution, and consumes 3200$\times$ less DRAM energy compared to a state‐of‐the‐art GPU‐only Floyd-Warshall.
\end{abstract}



\maketitle

\section{Introduction}

All-pairs shortest paths (APSP) looks for the shortest path between every pair of vertices in a graph. It is commonly found in routing, logistics, and network analysis problems ~~\cite{APSPA, APSPA1, APSPA2}.  APSP is often solved using dynamic programming via Floyd-Warshall (FW) algorithm, but its $O(N^3)$ complexity and massive data movement severely limit its performance on conventional CPU/GPU platforms. GPU acceleration has improved performance some by using blocked, or tiled, implementations ~\cite{CUDA}, the data-movement bottleneck remains a critical inhibitor. This "memory wall" necessitates a paradigm shift toward Processing-In-Memory (PIM) architectures. The blocked FW algorithm ~\cite{Blocked} is particularly well-suited for this approach, as its tiled structure exposes massive parallelism that aligns naturally with the inherent hierarchy of DRAM, offering a promising solution to the data-movement problem.

Despite significant GPU advancements---from initial CUDA ports~\cite{Harish2007,CUDA} to tiled and multi-GPU cluster approaches---performance remains bottlenecked by persistent data movement overheads~\cite{Scalable}, which dominate the energy profile of conventional architectures. The energy cost of this data transfer is pronounced that for a large-scale graph, a PIM approach that performs computation in memory can achieve an energy reduction compared to a state-of-the-art GPU baseline. This staggering inefficiency highlights the critical need for PIM and Process-Near-Memory(PNM) solutions.

PIM and PNM offer a compelling solution to the memory wall by integrating compute primitives directly within the memory hierarchy~\cite{PIM,PNM,Chi2016,AttAcc}. The primary distinction between these paradigms lies in the physical placement of the compute logic, which creates a fundamental trade-off between data locality and resource sharing. The PIM approach places logic directly within a memory bank, offering maximum energy efficiency and bandwidth by performing computation with minimal data movement. Conversely, the PNM approach places logic near the memory banks,  allowing it to be a shared resource for multiple bank-groups~\cite{Samsung}. While this incurs some local data movement, it provides a centralized and efficient point for tasks that require aggregating information from multiple sources.

While the structured memory access of dense graphs appears well-suited for the massive parallelism of modern GPUs, in practice, performance is severely bottlenecked by the memory wall. The $O(N^3)$ complexity of the FW algorithm translates to an overwhelming number of read-modify-write operations that saturate the memory bus. For instance, our analysis shows that this persistent data movement between a GPU's processing cores and its high-bandwidth memory is so costly that it accounts for the vast majority of the system's energy consumption, leading to a state-of-the-art NVIDIA A100 GPU requiring over 3,200$\times$ more energy than our PIM-based approach~. This bottleneck is architectural, not merely a matter of scale; even against a projected next-generation NVIDIA H100, our PIM architecture maintains a compelling 4.2$\times$ speedup, demonstrating that simply increasing bandwidth and compute resources in a conventional design yields diminishing returns for this memory-bound problem.

This highlights the need for in-memory computation. However, prior PIM/PNM graph accelerators have overwhelmingly focused on sparse graph processing~\cite{GraphPIM, GraphR}. Many of these solutions, such as GraphR~\cite{GraphR}, adopt a vertex programming model which spreads out vertices across different memory locations. This is efficient for managing the irregular, pointer-chasing memory access patterns of sparse workloads (e.g., Breadth-First Search), but it is fundamentally inefficient for the dense, all-to-all workload of APSP. While The prior sparse-graph PIM accelerators adopt vertex programming mode which spreads out vertices across different memory locations. However, when processing fully dense graph applications (e.g., ASAP), the communication costs of transferring vertex data between different memory components become prohibitively large. This renders these sparse-optimized designs ill-suited for dense APSP, creating a key need for an architecture that addresses the unique challenges of dense, matrix-based graph algorithms.

Our work addresses this gap by proposing PIM-FW, the first DRAM-based architecture co-designed to accelerate blocked FW algorithm. The core insight of our work is a pragmatic hardware-software co-design that divides the algorithm's tasks based on their computational characteristics. On the hardware side, we propose a hybrid architecture: for the dominant and massively parallel $O(N^3)$ \texttt{min-plus} updates, we employ a pure PIM approach by embedding massive arrays of compact, bit-serial processing elements, which we term Bank PEs (BPEs), within each DRAM bank. This design choice performs computation , localizing the bulk of the operations to maximize parallelism and virtually eliminate the data-movement bottleneck inherent in GPU-based approaches. Conversely, for communication-intensive reduction tasks, we adopt a PNM model, utilizing a shared engine at the channel level, termed Channel PE (CPE), to efficiently aggregate results from across the bank-groups. The core BPE and CPE processing units share the same fundamental design; their primary distinction is their physical placement: BPEs are distributed in-bank for maximum locality, while the CPE is centralized for efficient cross-bank-group communication. On the software side, we complement this hybrid hardware with a novel interleaved data mapping policy that ensures superior load balancing~. Through cycle-accurate simulation, this co-design demonstrates remarkable results: PIM-FW achieves an 18.7$\times$ speedup and an over 3,200$\times$ energy reduction compared to a state-of-the-art NVIDIA A100 GPU.

\section{PIM-FW Design for APSP Acceleration}

\begin{figure}[!htbp]
    \centering
    \includegraphics[width=\linewidth]{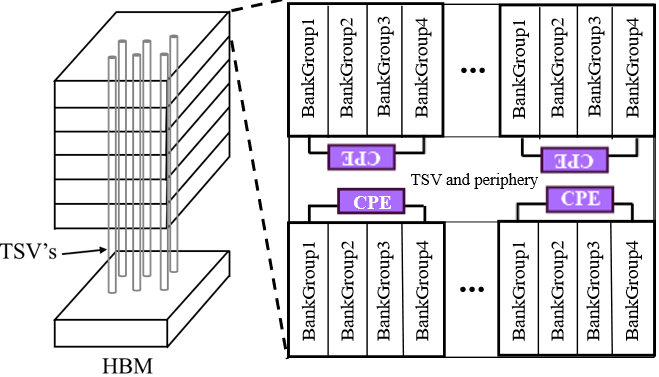}
    \caption{The PNM architecture of PIM-FW, showing shared Channel PEs (CPEs) for global reduction. Its processing unit design is identical to the Bank PE (BPE), detailed later.}
    \label{fig:overall}
\end{figure}

Our architecture, shown in Fig.~\ref{fig:overall}, is built upon High-Bandwidth Memory (HBM3)~\cite{HBM3}, a 3D-stacked DRAM technology that provides massive bandwidth and enables fine-grained parallelism~\cite{TransPIM2023}. Its key architectural features, such as a wide data interface interconnected by Through-Silicon Vias (TSVs) and a deep hierarchy of independent channels and bank-groups, are fundamental to our design. This structure allows for concurrent operations across multiple channels and bank-groups, providing the necessary foundation for our accelerator.
We propose PIM-FW that accelerates the blocked FW algorithm through a novel dataflow and a hybrid PIM architecture. We chose blocked FW due to its high parallelism, as discussed below in Section~\ref{blockedFW}.

The PIM-FW operation begins with a high-throughput bulk transfer of the tile-major ordered matrix into the HBM3 stack, leveraging the wide 1024-bit TSV interface.  Building on this hierarchy, we integrate CPE at the channel level, responsible for global reduction tasks such as comparing minimum values across different bank-groups. The overall execution is orchestrated by a host CPU for initial data loading, which then delegates control to the main memory controller. This controller issues specialized commands to manage the algorithm's phased execution, and its associated latency and energy overhead are fully accounted for in our cycle-accurate simulation. The detailed hardware implementation of our PIM-FW architecture is presented in Section~\ref{PIM-FW}. The specific data alignment and scheduling strategies designed to maximize parallelism are further detailed in Section~\ref{mapping}.

\subsection{Traditional and Blocked Floyd–Warshall Algorithms}
\label{blockedFW}

The All-Pairs Shortest-Path (APSP) problem aims to find the shortest path distance for every pair of vertices \((u,v) \in V\), with the solution being an \(N \times N\) distance matrix \(D\). The canonical FW algorithm is a fundamental method for solving this problem~\cite{AttAcc,GPUAPSP}. The algorithm operates on a dense distance matrix, iteratively updating every path using the relaxation ``D[i,j] = min(D[i,j], D[i,k] + D[k,j])''\cite{Floyd1962}. The fundamental computation at the heart of this process is the min-plus operation.However, the naïve implementation's $O(N^3)$ complexity and poor data locality create a severe \textbf{data-movement bottleneck} on conventional CPU and GPU platforms~\cite{Harish2007,Scalable}. This bottleneck arises because for each \texttt{min-plus} operation in the innermost loop, the processor must read three values (`D[i,j]`, `D[i,k]`, and `D[k,j]`) from and write one value back to main memory. For a graph of size N, this results in approximately $4 \times N^3$ memory access operations. When N is large, the time and energy spent waiting for data to travel between the processor and main memory far exceed that of the actual computation, making memory bandwidth---not computational power---the limiting factor for performance.

To mitigate this, the blocked FW algorithm partitions the matrix into smaller \(B \times B\) tiles. This tiling strategy enhances data locality by allowing data for multiple updates to be held in fast on-chip memory~\cite{Blocked speed,constraints}. Furthermore, it exposes massive parallelism between independent tile updates, making it a foundational strategy for high-performance implementations and particularly well-suited for the PIM architectures we explore in this work. The procedure is detailed in Algorithm~\ref{alg:blocked_fw}.

\begin{algorithm}[h!]
\caption{Blocked Floyd-Warshall Algorithm}
\label{alg:blocked_fw}
\begin{algorithmic}[1]
\Require Adjacency matrix $D$ of size $N \times N$, block size $B$
\Ensure Matrix $D$ with all-pairs shortest paths

\State Let $M = N/B$ and let $A_{ij}$ be the $(i, j)$-th block of $D$

\For{$k \gets 0$ to $M-1$}
    \State \Comment{Phase 1: Update pivot block}
    \State $A_{kk} \gets \text{Floyd-Warshall}(A_{kk})$

    \State \Comment{Phase 2: Update pivot row and column}
    \State For each $j \in \{0..M-1\}, j \neq k$: \quad $A_{kj} \gets \min(A_{kj}, A_{kk} \oplus A_{kj})$
    \State For each $i \in \{0..M-1\}, i \neq k$: \quad $A_{ik} \gets \min(A_{ik}, A_{ik} \oplus A_{kk})$

    \State \Comment{Phase 3: Update remaining blocks}
    \State For each $i,j \in \{0..M-1\}, i \neq k, j \neq k$: \quad $A_{ij} \gets \min(A_{ij}, A_{ik} \oplus A_{kj})$
\EndFor

\State \Return $D$
\end{algorithmic}
\end{algorithm}

\subsection{PIM-FW: Hardware for APSP Acceleration}
\label{PIM-FW}
\begin{figure}[!htbp]
    \centering
    \includegraphics[width=\linewidth]{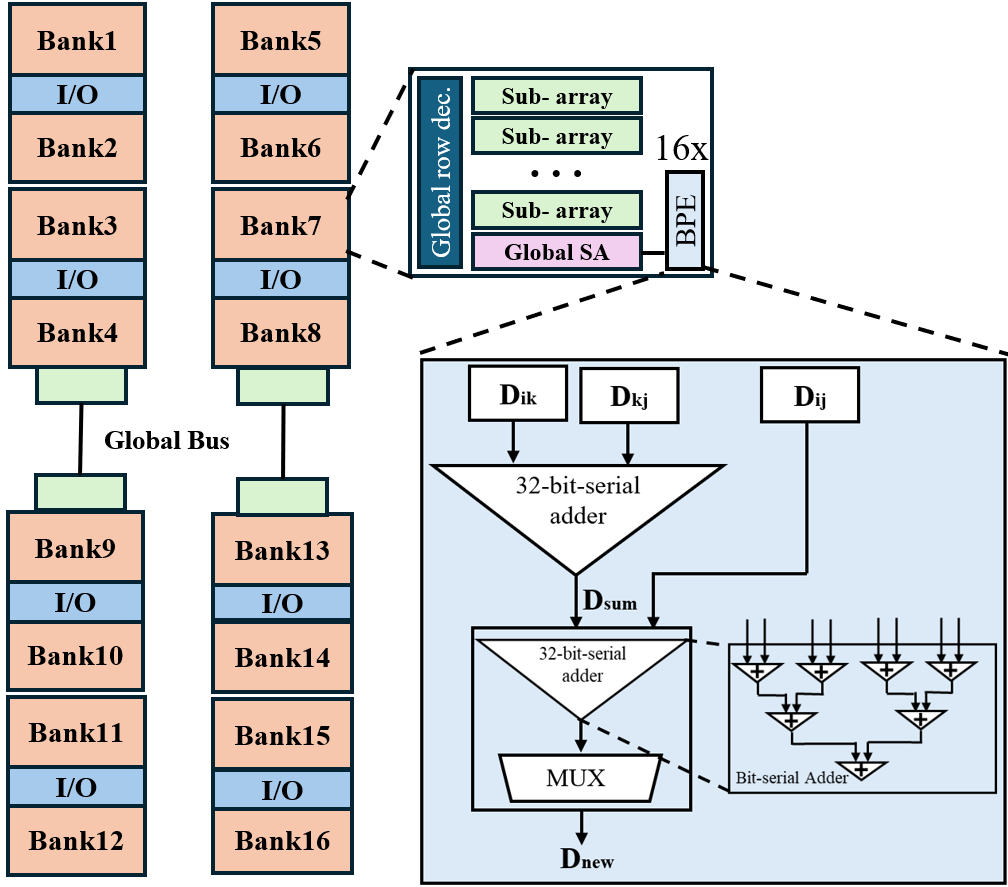}
    \caption{The composition of a single bank-group and the internal logic of a single PIM PE (BPE), which uses a bit-serial adder and a MUX to efficiently perform the \texttt{min} operation.}
    \label{fig:HBM}
\end{figure}
The in-memory processing of our architecture employs a sophisticated hybrid model, strategically combining the strengths of PIM and PNM to optimize different stages of the blocked FW algorithm based on their computational characteristics.

\textbf{PIM(BPE) for core \texttt{min-plus} updates:} The bulk of the FW algorithm is dominated by $O(N^3)$ \texttt{min-plus} computations, which are massively parallel and data-local. To tackle this, we adopt a PIM approach: as shown in Fig.~\ref{fig:HBM}, each DRAM bank is equipped with an array of 16 BPEs. This design minimizing data movement for the core workload and maximizing energy efficiency and parallelism.

\textbf{PNM(CPE) for communication-intensive reductions:} Conversely, stages such as finding a minimum value across different bank-groups for a global reduction are inherently communication-intensive. For such tasks, a centralized, more capable unit is more efficient. We therefore employ a CPE model Fig.~\ref{fig:overall}, utilizing a shared engine at the channel level. This CPE engine is responsible for efficiently aggregating results from its constituent bank-groups, avoiding the significant overhead of coordinating complex cross-bank communication among thousands of simple in-bank BPEs.

This pragmatic division of labor allows us to match each stage of the algorithm to its most efficient processing paradigm, which is the key novelty of our PIM-FW architecture.

The interaction between a BPE and the DRAM components for a single `min-plus` operation follows a precise sequence. The process begins when the DRAM controller opens a wordline and latches the row's 8192 bits of data into the row buffer via the bitlines and sense-amplifiers. Once the data is latched, each of the 16 BPEs within the bank reads its required 32-bit operand from a dedicated 512-bit slice of the row buffer's internal datalines. To support the `min-plus` operation, as detailed in Fig. \ref{fig:HBM}, the BPE logic first uses a bit-serial adder to compute the sum ($D_{sum} = D_{ik} + D_{kj}$) and then uses a bit-serial adder and multiplexer to efficiently perform the comparison and select the minimum value. The final 32-bit result (\(D_{\text{new}}\)) is then written back to its corresponding slice in the row buffer, overwriting the old data. 

Conversely, for the communication-intensive reduction phase, we utilize an CPE paradigm. A global processing engine at the channel-level controller collects the local minima from its constituent bank-groups via the on-die interconnects. It uses an internal comparison tree to find the channel's overall minimum in approximately 5-10 cycles. The final global minimum is then found by a reduction across all channel-level results at the main memory controller. This hybrid approach uses the most efficient paradigm for each specific task.

\subsection{PIM-FW: APSP Data Mapping and Scheduling}

\subsubsection{Data Mapping}
\label{mapping}

\begin{figure}[!htbp]
    \centering
    \includegraphics[width=\linewidth]{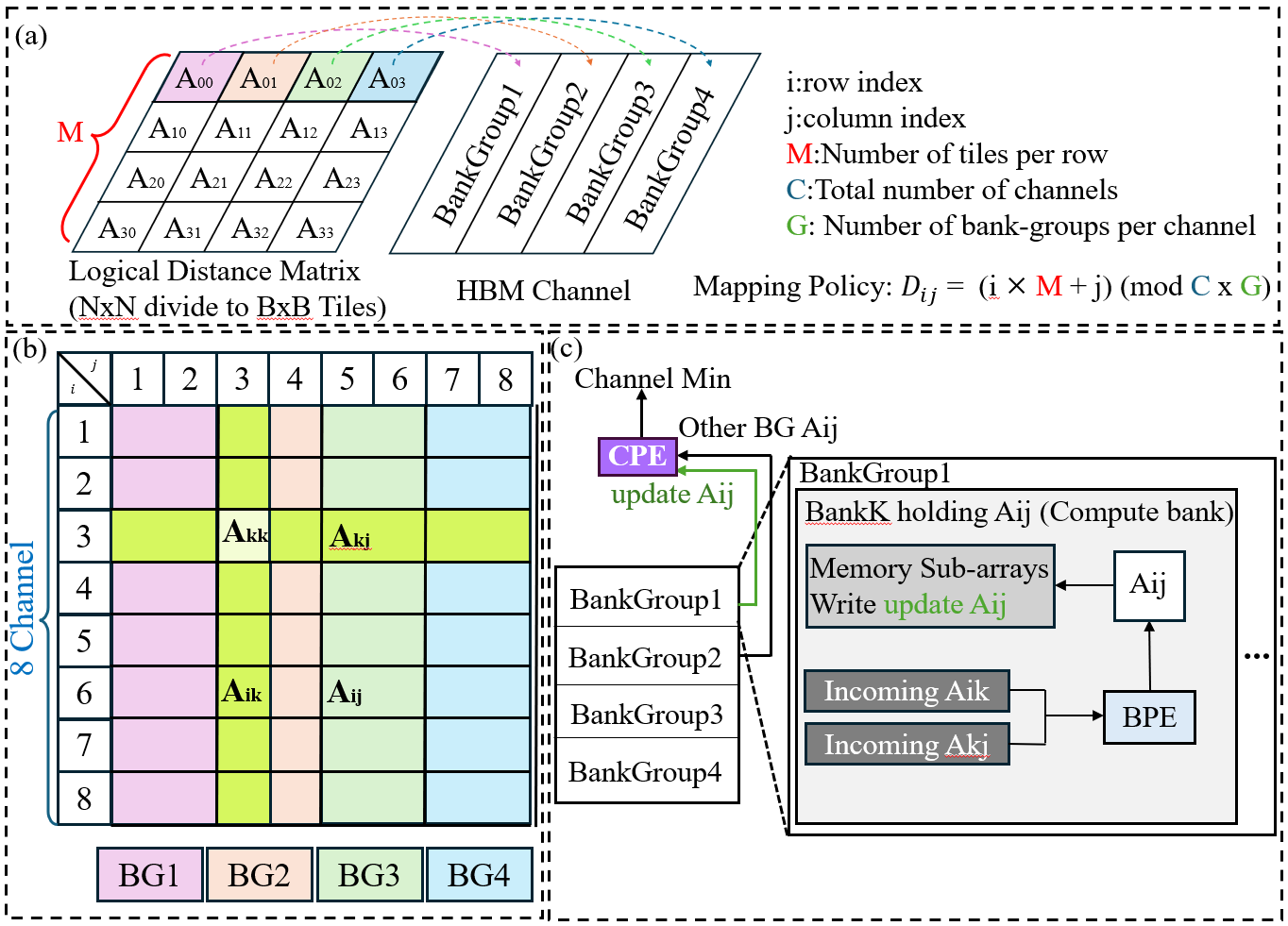}
    \caption{(a)Mapping policy (b)Dependency of blocked Floyd–Warshall algorithm: bank‐group(BG) view showing phase 1 (pivot), phase 2 (row/column), and phase 3 (independent).(c)PIM-FW Execution Flow for a Parallel Update. }
    \label{fig:tile_bank_phases}
\end{figure}

To efficiently execute the blocked FW algorithm on our PIM architecture, we employ a compiler-driven \cite{OptiPIM, CIM}, co-designed dataflow centered on a novel interleaved tile Mapping policy. The dataflow partitions the $N \times N$ distance matrix into $B \times B$ tiles and distributes them across the HBM3's 32 bank-groups via a static, interleaved mapping policy. This policy maps a logical tile index $(i, j)$ to a physical bank-group ID using the following formula shown in Fig.~\ref{fig:tile_bank_phases}(a):
$\text{bank\_group\_id} = (i \times M + j) \pmod{C \times G}$,  where (\(i, j\)) are the tile's logical coordinates, \(M\) is the number of tiles per row, \(C\) is the total number of channels, and \(G\) is the number of bank-groups per channel.  This mapping maximizes parallelism for the underlying hybrid hardware, which leverages massively parallel BPEs (PIM) for core \texttt{min-plus} updates and CPEs (PNM) for global reduction tasks.

Fig.~\ref{fig:tile_bank_phases} (b) further clarifies the multi‐level dependency structure of our blocked FW implementation.  The execution of each pivot iteration is governed by a three-phase data dependency. Phase 1 first updates the pivot tile \(A_{kk}\) in isolation. In phase 2, the results from \(A_{kk}\) are broadcast to all tiles in the pivot row (\(A_{kj}\)) and column (\(A_{ik}\)), which can then be processed in parallel. Finally, phase 3 updates the remaining off-diagonal tiles (\(A_{ij}\)) in a parallel wavefront, using data broadcast from the tiles computed in phase 2 — specifically, rows from \(A_{ik}\) and columns from \(A_{kj}\). This entire flow is enabled by the HBM3's wide 1024-bit TSV interface, which can broadcast entire B-element pivot vectors to all target bank-groups in just a few cycles, unlocking the parallelism in both phases 2 and 3. Within each phase, all white (off‐diagonal) entries across different sub‐blocks are free of inter‐block dependencies and therefore can be executed fully in parallel.

Fig.~\ref{fig:tile_bank_phases} (c) shows the execution of a parallel update in our hybrid architecture is detailed. Pivot data from source bank-groups (holding $A_{ik}$ and $A_{kj}$) is broadcast to the compute bank-group holding $A_{ij}$. Inside the compute bank, BPEs perform the \texttt{min-plus} operation, while the CPE handles communication-intensive reduction tasks.

\subsubsection{Exploiting Coarse and Fine-Grained Parallelism}
Our architecture and mapping policy are co-designed to exploit parallelism at two distinct granularities:
\begin{itemize}[leftmargin=*]
    \item \textbf{Coarse-grained parallelism:} At the highest level, our design leverages the independence of HBM3 channels. Since our mapping policy distributes tiles across all channels, multiple independent tile updates (e.g., during phase 3 wavefronts) can occur simultaneously in different channels, each with its own controller and data path. This channel-level parallelism is essential for achieving high throughput on the overall problem.

    \item \textbf{Fine-grained parallelism:} Within each bank-group, our design achieves fine-grained parallelism via the in-bank PE arrays. A single bank-group contains a total of 256 PEs (16 banks $\times$ 16 PEs/bank). This number is deliberately chosen to match our tile dimension ($B=256$). This allows an entire 256-element row of a tile to be processed in a single computational pass, with each PE handling one element. This internal parallelism is what allows each tile-level update to be executed with efficiency.
\end{itemize}

By combining this mapping policy with a hardware design that supports both coarse- and fine-grained parallelism, we can fully exploit the dependencies of the blocked FW algorithm to achieve significant performance gains.

\subsubsection{Data Scheduling}
\begin{figure}[!htbp]
    \centering
    \includegraphics[width=\linewidth]{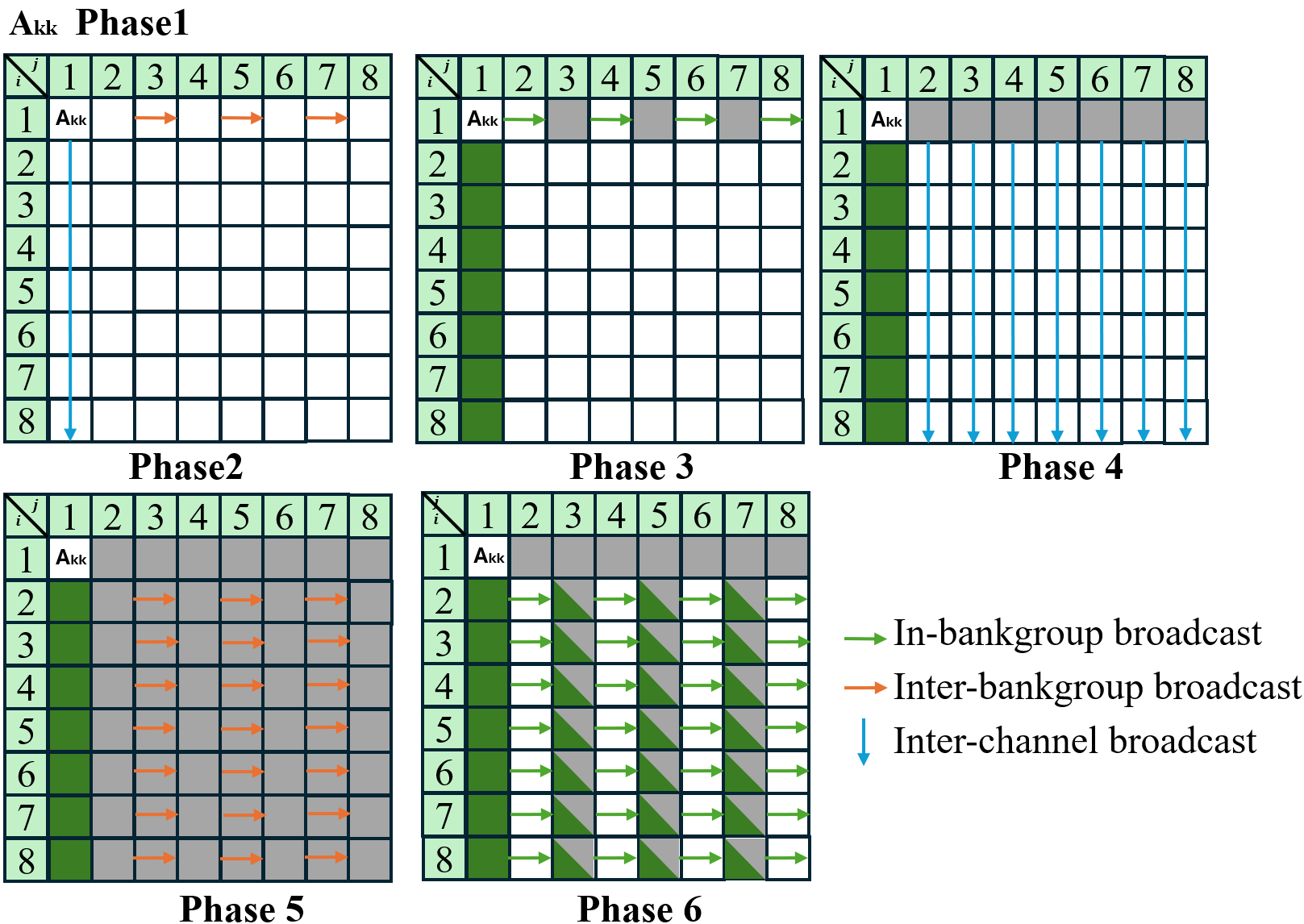}
    \caption{Blocked Floyd-Warshall data movement and scheduling on a logical 4x8 tile grid.}
    \label{fig:movement}
\end{figure}

To efficiently map the blocked Floyd-Warshall algorithm onto our hardware, we devised a multi-phase schedule, illustrated in Fig.~\ref{fig:movement}. This schedule is a necessary adaptation to impose the algorithm's logical 8x8 tile grid onto the physical 4x8 HBM3 organization, which consists of 4 bank-groups per channel across 8 independent channels. The process begins in phase 1 with the self-contained update of the pivot block, $A_{kk}$. In phase 2, the updated pivot data is broadcast to all other tiles in its corresponding row and column. Due to the architecture, broadcasting is parallel across different channels but sequential within a single channel. This constraint means the inter-bankgroup broadcast (represented by the orange arrow) requires three broadcast cycles. In contrast, the in-bankgroup broadcast (represented by the green arrow) and the inter-channel broadcast (represented by the blue arrow) are each completed in a single cycle. Following the broadcast, phases 3 and 4 show the update of the tiles in the pivot column and row, respectively. Finally, phases 5 and 6 show the remaining off-diagonal tiles being updated in a wavefront pattern.

\section{EVALUATION}
\subsection{Experimental Setup}


To evaluate PIM-FW, we use the Ramulator2-based\cite{Ramulator} cycle-accurate model to simulate our HBM3-based design. Our implementation targets a standard HBM3 stack with the following configuration: 8 channels per ie, each channel comprising 16 banks organized into 4 bank-groups; a 4-Hi die stack with 32k rows per bank, 1 KB row size and 512×512-cell subarrays; a 256-bit wide DQ bus yielding 1024 TSV lanes and a peak bandwidth of 256 GB/s (2 Gbps/pin at 1.2 V). The DRAM timing parameters used in our simulation are \(t_{RC}=30\) ns, \(t_{RCD}=8\) ns, \(t_{RAS}=24\) ns,  \(t_{RRD}=2\) ns, \(t_{WR}=12\) ns, \(t_{CCDS}=2\) ns, and \(t_{CCDL}=4\) ns,the detailed parameters for our simulated HBM3 stack and PIM engines are summarized in Table~\ref{fig:HBM3 spec}. This hardware configuration, particularly the 32 total bank-groups, defines our maximum parallelism ($N_{\text{max,par}}$) and thus limits the maximum graph size to $N=8,192$ (where $M \le 16$ and $B=512$), which we use as the largest dataset in our evaluation. We estimate the HBM area using CACTI-3DD \cite{CACTI} at 22nm technology node. The area and power of our PEs are obtained by synthesizing the RTL using a 65nm library, with the results subsequently scaled to 22nm. For the GPU baseline, we report the total energy consumed during execution, measured using the \texttt{nvidia-smi} utility. This provides a comprehensive measure of the entire board's power draw, ensuring a fair system-level comparison.

\textbf{Datasets}:
We evaluate PIM-FW across three real-world graphs from the SNAP dataset \cite{SNAP} and OpenStreetMap \cite{OSM}: ca-GrQc ($N=5,242$), p2p-Gnutella08 ($N=6,301$) , and OpenStreetMap ($N=8,192$). Our evaluation uses inherently dense graphs; to align with the FW algorithm's requirements, we represent non-existent edges with infinite weight. While specialized algorithms exist for sparse graphs~\cite{APSP}, the dense FW approach is fundamental for problems where graphs are dense or edge weights change frequently, making re-computation on a dense matrix more practical.

\textbf{GPU baselines}: 
Our primary baseline is a highly-optimized blocked FW implementation on an NVIDIA A100 (SXM4, 40GB) GPU. Furthermore, we provide a performance projection for the NVIDIA H100 GPU\cite{H100}.  This projection is a conservative estimate based on the H100's documented architectural advantages, such as significantly higher memory bandwidth and more streaming multiprocessors.

\begin{table}[!htbp]
    \centering
    \caption{\textbf{parameter for the APSP in HBM3}}
    \label{fig:HBM3 spec}
    \includegraphics[width=1 \linewidth]{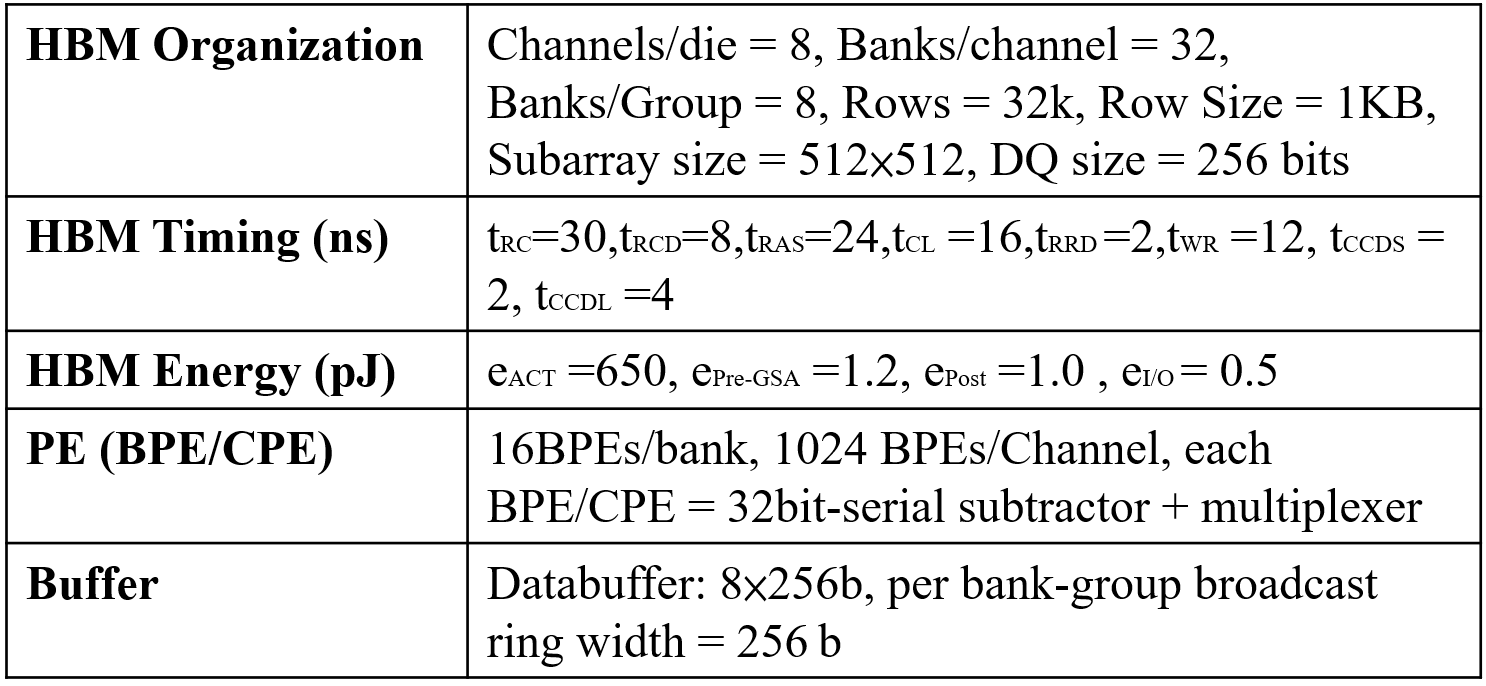}
\end{table}

\subsection{Performance on Real-World Graph Datasets}
To validate the performance of PIM-FW on practical applications, we evaluate our design on three large-scale, real-world datasets, specifically chosen to represent diverse network topologies: the \texttt{ca-GrQc} collaboration network (\(N=5,242\)), the \texttt{p2p-Gnutella08} peer-to-peer network (\(N=6,301\)) \cite{{SNAP},OSM}, and the OpenStreetMap road network (\(N=8,192\)).  This dataset is specifically chosen to evaluate the boundaries of our design, as $N_{\text{max,par}}=8,192$, which, in turn, is determined by the design's parallelism constraints:  our HBM3 stack provides 32 bank-groups, and since the blocked FW algorithm's wavefront loads $2M$ tiles simultaneously, the maximum $M$ is limited to 16 ($2M \le 32$). Combined with a block parameter $B=512$, this yields the $N=8,192$ limit.

\begin{figure}[!htbp]
    \centering
    \includegraphics[width=\columnwidth]{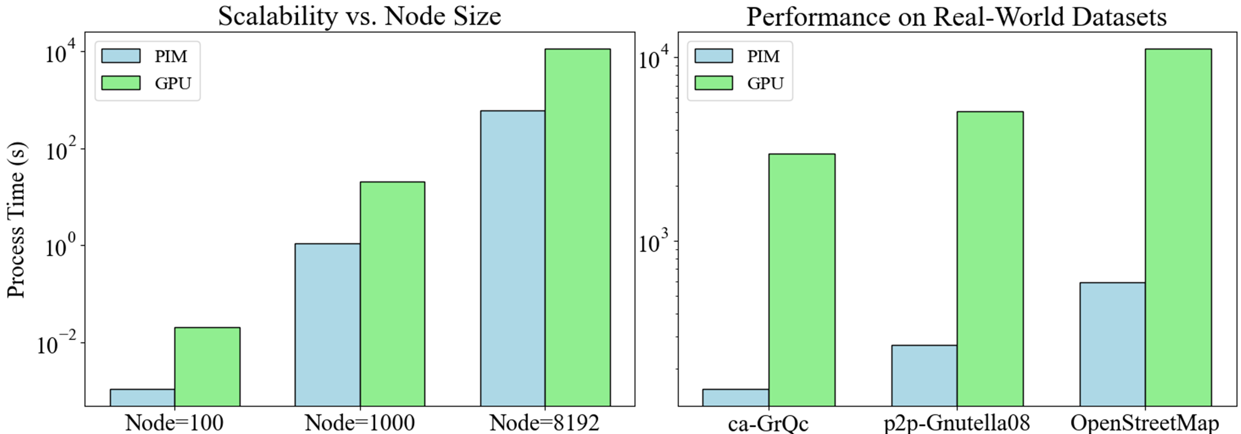}
    \caption{Performance comparison of the PIM-based architecture and a baseline GPU.}
    \label{fig:dataset}
\end{figure}

The selection of these distinct graph structures is intended to demonstrate the broad applicability and robustness of our architecture across varied computational workloads, proving its relevance to real-world scenarios.
In our evaluation, these inherently sparse graphs are treated as complete graphs where non-existent edges are assigned an infinite weight to align with the requirements of a dense APSP accelerator.
Performance projections, extrapolated from our measured results, are presented in Fig.~\ref{fig:dataset}.
Notably, for the all dataset, by implementing our mapping policy, we observed a ~3.2$\times$ speedup. With hardware implementing it yielded an additional ~5.8$\times$ speedup. Our PIM-based architecture achieves a significant speedup of approximately 18.7$\times$ over the state-of-the-art NVIDIA A100 GPU.
This consistent high performance across topologically different datasets underscores the capability of our PIM-based design to efficiently manage graph analytics workloads.

We assess our architecture's scalability by projecting performance across various graph sizes based on the Floyd-Warshall algorithm's $O(N^3)$ complexity. Using our measured runtimes on the N=8,192 dataset (592s for PIM-FW vs. 11,097s for the A100 GPU) as a baseline, our projections confirm that PIM-FW's significant performance advantage is consistently maintained across all scales. For instance, at N=1,000, PIM-FW is projected to require just 1.08 seconds, while the A100 GPU would take 20.2 seconds, demonstrating robust scalability for real-world applications.


\subsection{Performance and Energy Efficiency Results}



\begin{figure}[!htbp]
    \centering
    \includegraphics[width=0.8\linewidth]{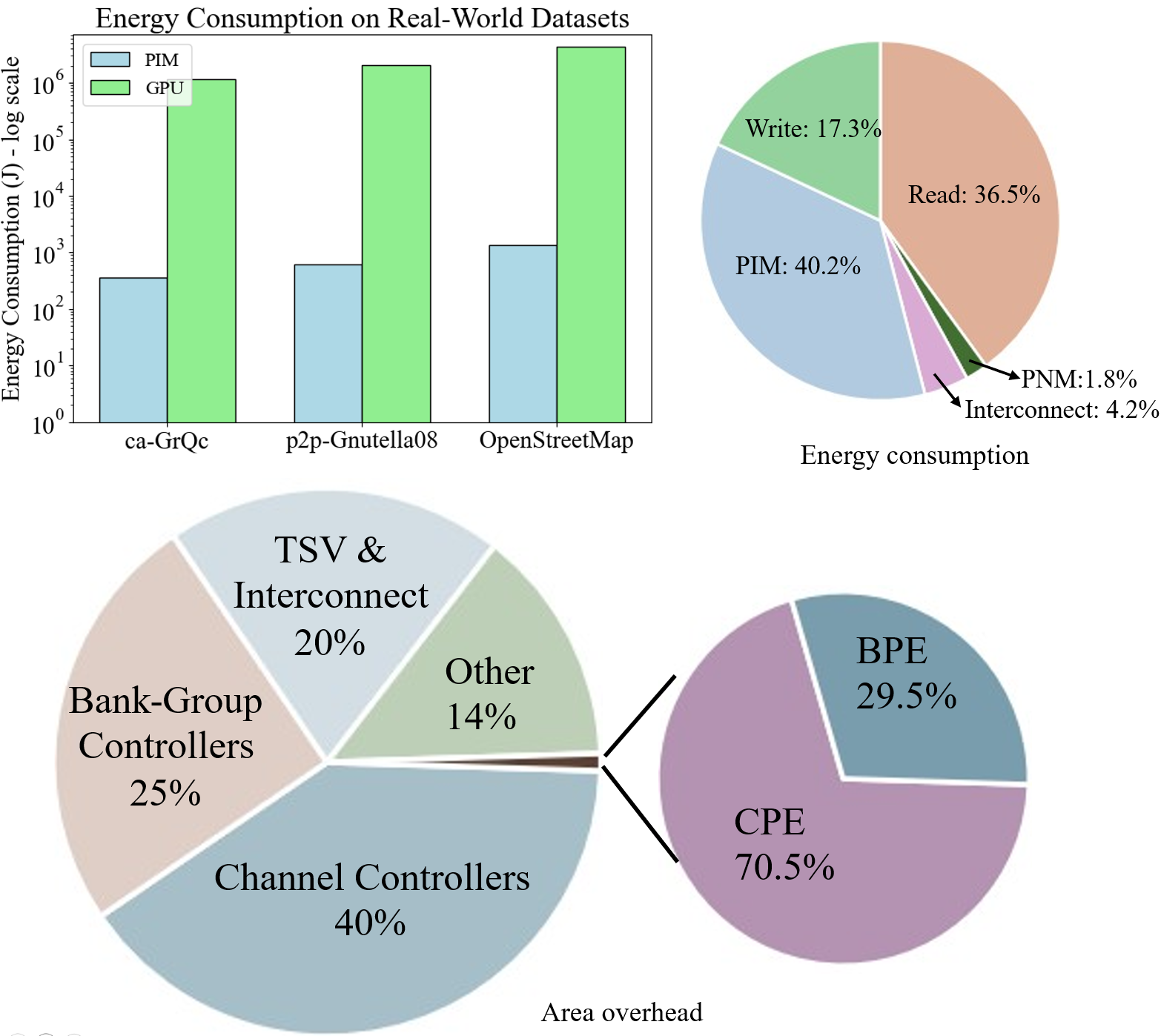}
    \caption{Total energy consumption (upper left) for the same datasets.The energy consumption (upper right) is dominated by DRAM read and write and PIM computations for the PIM-FW. The area analysis (below) for the PIM-FW.}
    \label{fig:energyarea}
\end{figure}

For a large-scale graph of $N=8,192$, the total execution time of our proposed architecture is 592 seconds, while the NVIDIA A100 GPU takes 11,097 seconds to complete the same task. This represents a significant speedup of 18.7x over the GPU baseline. As shown in Fig. \ref{fig:energyarea}, this performance advantage holds across all tested real-world datasets.

The energy efficiency of our design is even more pronounced. Across all datasets, the energy required by the GPU is several orders of magnitude higher than that of our PIM design, corresponding to a normalized energy reduction factor of over 3,200x. This exceptional efficiency stems directly from our PIM-based approach, which nearly eliminates the costly data movement between processors and memory that dominates the energy profile of conventional architectures.

While our primary baseline is the A100, we also provide a performance projection for the NVIDIA H100 GPU-Estimate. This projection is an estimate based on the H100's documented architectural advantages, including significantly higher memory bandwidth and more streaming multiprocessors. Based on these factors, we conservatively estimate the H100 to be approximately 4.5x faster than the A100 \cite{H100} on this workload. Even against this projected baseline, PIM-FW would maintain a compelling 4.2x speedup.


\subsection{Energy and Area Overhead Analysis}

The PIM-FW architecture demonstrates exceptional efficiency in both energy consumption and silicon area overhead, validating the feasibility of our approach. Cycle-accurate simulations reveal over 3,200$\times$ energy reduction against state-of-the-art GPUs, with BPEs consuming the majority of compute energy for the core $O(N^3)$ updates. This validates PIM-FW's feasibility and superior efficiency, leveraging the architectural alignment of blocked FW with DRAM's physical hierarchy. The area overhead shows our added PIM logic (BPE+CPE) is a negligible fraction of the total HBM3 logic die. The majority of the die area is consumed by standard HBM components, including channel/bank-group controllers, TSV interconnects, and other peripheral circuits for functions like power delivery. The vast majority of a typical HBM3 logic die is occupied by standard components. Our proposed PIM logic constitutes less than 1\% of the total die area. Within this added PIM logic, the CPE and their interconnect logic account for \textasciitilde76\% of the area, while the massive array of 8,192 highly-compact PIM BPEs accounts for the remaining \textasciitilde24\%.


\subsection{Impact of Channel and BPE Level Parallelism}

The architecture's coarse-grained parallelism scales near-linearly with the number of independent HBM3 channels. Since each channel operates independently, a higher channel count directly enables more concurrent tile updates and parallel data distribution via the TSV bus, mitigating serialization bottlenecks during broadcast-heavy phases. As shown in the provided  Fig.~\ref{fig:energyarea} (upper right), scaling from 4 to 32 channels reduces the total process time by approximately 8x. This result underscores the design's scalability with future HBM standards that may feature more channels.

Fine-grained performance is dictated by BPE-level parallelism within each bank-group. Our baseline design is optimized by matching the number of BPEs per bank-group (256) to the tile width (\(B=256\)), which allows a 256-element tile row to be processed in a single computational pass. Halving the number of BPEs to 128 doubles the required passes, increasing execution time by \textasciitilde1.45x, while quartering them to 64 increases it by \textasciitilde2.35x. Conversely, increasing the BPEs to 512 provides no additional speedup, as the extra hardware would sit idle. Therefore, 256 BPEs per bank-group represents the optimal design point, maximizing computational throughput without wasting silicon resources. We could continue to increase the BPE count, which might allow processing larger tile widths dataset, but for the overall acceleration of the current workload, this would have no significant effect.

\begin{figure}[!htbp]
    \centering
    \includegraphics[width=\linewidth]{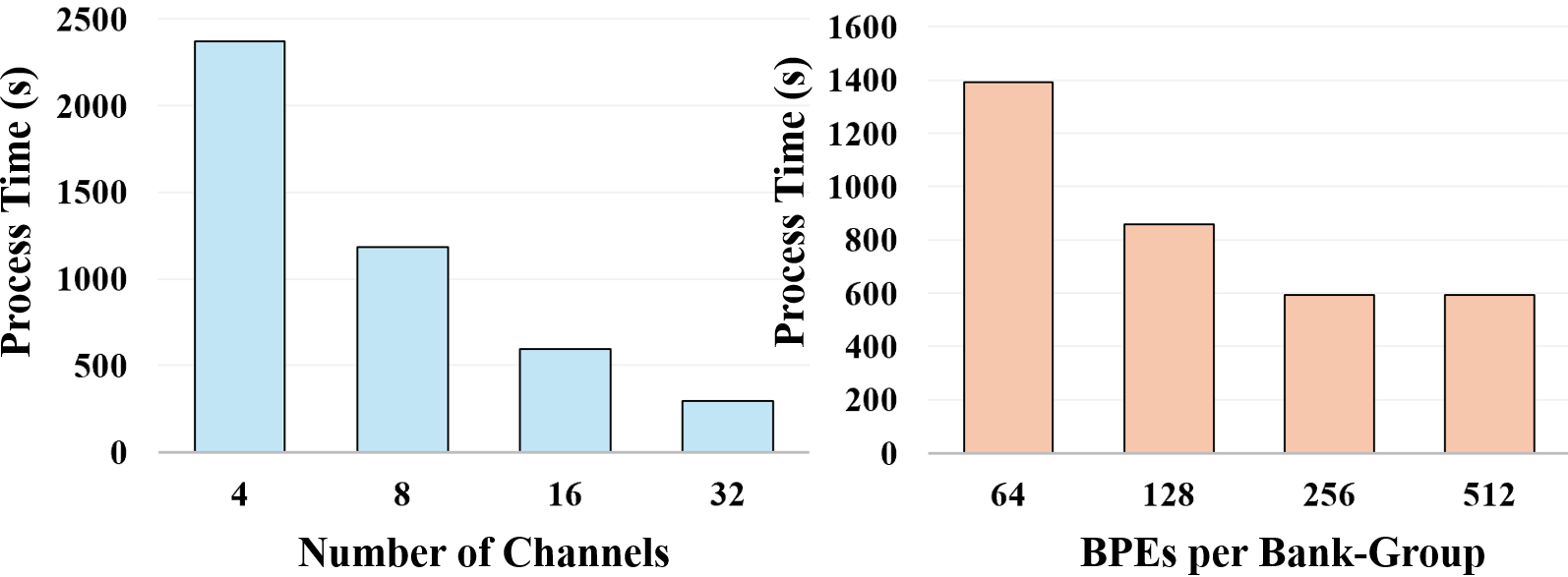}
    \caption{The result for different BPEs and Channels at N=8192 process time}
    \label{fig:compare}
\end{figure}

We also evaluate PIM-FW against other heterogeneous APSP accelerators. For instance, recent work on an FPGA-based solution ~\cite{FPGA} for N=8192 graphs reported a 137$\times$ speedup, but this was compared to a CPU-only baseline, with authors noting their performance was near-GPU levels. In sharp contrast, PIM-FW achieves its 18.7$\times$ speedup compared to the state-of-the-art NVIDIA A100 GPU baseline~\cite{A100}. Given that the FPGA solution only approaches GPU performance, therefore PIM-FW is at least 18.7$\times$ faster than the FPGA solution ~\cite{FPGA}. Interestingly, both approaches converge on a heterogeneous solution to solve APSP: the FPGA work utilizes a CPU-FPGA co-design, while our PIM-FW implements a hybrid PIM/PNM architecture (BPE/CPE) to manage the computation and the communication bottleneck.

\section{Conclusion}
In this paper, we addressed the significant data movement bottleneck of the blocked Floyd-Warshall algorithm by introducing PIM-FW, a novel hardware-software co-design for dense APSP graphs. The key to our approach is a pragmatic hybrid architecture: massively parallel in-bank Processing Elements (PIM) handle the core $O(N^3)$ computations, while a more powerful channel-level engine (PNM) manages communication-intensive reduction tasks, all complemented by a static, interleaved data mapping policy. OurRamulator2-based\cite{Ramulator} cycle-accurate evaluation demonstrates the efficacy of this co-design, showing that PIM-FW achieves a remarkable 18.7$\times$ speedup and over 3,200$\times$ energy reduction compared to a highly-optimized NVIDIA A100 GPU baseline, while maintaining a compelling 4.2$\times$ speedup against a projected next-generation NVIDIA H100. Furthermore, our analysis indicates that while the current BPE count is optimal for our baseline, increasing BPE parallelism could enable the processing of even larger datasets by supporting wider computational tiles ($B$). These results show that our co-designed PIM approach, which closely couples computation with a tailored data placement strategy, is a powerful and a viable solution for accelerating memory-intensive dynamic programming workloads.



\end{document}